\documentclass[final,5p,times,twocolumn]{elsarticle}

\usepackage{amsmath,amssymb,amsfonts}
\usepackage{graphicx}
\usepackage{bm}
\usepackage{booktabs}
\usepackage{array}
\usepackage{xcolor}

\usepackage{cuted}
\newenvironment{widetext}{\begin{strip}\centering}{\end{strip}}

\newcommand{\sqs}{\sqrt{s}}
\newcommand{\Xic}{\Xi_c}
\newcommand{\Xicp}{\Xi_c'}
\newcommand{\Xicst}{\Xi_c^{*}}
\newcommand{\Lamc}{\Lambda_c}

\journal{Physics Letters B}

\begin{document}

\begin{frontmatter}

\title{Five-flavor $udsc\bar{b}$ molecular pentaquarks from heavy-quark and local hidden gauge symmetries}

\author[a]{Ratirat Suntharawirat}
\ead{ratirat_su@kkumail.com}
\author[a]{Nongnaphat Ponkhuha}
\ead{Nongnapat.po@kkumail.com}
\author[a]{Daris Samart\corref{cor}}
\cortext[cor]{Corresponding author.}
\ead{darisa@kku.ac.th}
\address[a]{Khon Kaen Particle Physics and Cosmology Theory Group (KKPaCT), 
Department of Physics, Faculty of Science, 
Khon Kaen University, 
123 Mitraphap Rd., Khon Kaen 40002, Thailand}

\begin{abstract}
We study a family of genuinely exotic five-flavor molecular pentaquark states containing the five quark flavors $u,d,s,c,b$ that form experimentally accessible hadrons. We construct the meson-baryon interaction from the local hidden gauge symmetry combined with heavy-quark spin symmetry, following the chiral unitary description that reproduces the LHCb hidden-charm strange pentaquarks. Heavy-quark flavor symmetry allows us to obtain the $udsc\bar{b}$ sector by replacing the anti-charm quark in the meson with an anti-bottom quark while keeping the charm quark in the baryon. As a result, we obtain ten threshold-associated isoscalar poles with $J^P=1/2^-,3/2^-,5/2^-$ in the range $7.72$ to $7.96$ GeV. They are narrow and organized into heavy-quark spin multiplets with predicted near-degeneracies. There are four states that overlap with the earlier two-sector study in the literature to about $2$ MeV, which shows that the separate-sector treatment is recovered as a limit of this work. Moreover, we also identify two additional, more deeply bound $B_s\Lambda_c$ and $B_s^{*}\Lambda_c$ poles generated by strong inter-channel coupling because these poles are farther from their dominated thresholds. This is an interesting signal that can be searched for at LHCb in the $B_c\Lambda$ and $B_s^{*}\Lambda_c$ invariant mass spectra. 
\end{abstract}

\begin{keyword}
exotic hadrons \sep pentaquark \sep hadronic molecule \sep heavy-quark symmetry \sep coupled channels
\end{keyword}

\end{frontmatter}

\section{Introduction}

The observation by the LHCb collaboration of the $P_c(4312)$, $P_c(4440)$ and $P_c(4457)$ peaks in the $J/\psi p$ invariant mass distribution \cite{Aaij2015,Aaij2019} confirmed a decade-old prediction that $S$-wave $\bar{D}^{*}\Sigma_c$ systems should bind to hidden-charm pentaquarks \cite{Wu2010,Wu2011}. The masses of these states are located just below the $\bar{D}\Sigma_c$ and $\bar{D}^{*}\Sigma_c$ thresholds, which supports the molecular interpretation \cite{Chen2016,Guo2018}. The model naturally reproduces both the narrow widths of these three peaks and their alignment with the relevant thresholds, treating them as loosely bound \(\bar{D}^{*}\Sigma_c\) molecules \cite{ChenChenZhu2019,RChen2019,Karliner2015}, a picture that has been developed in many complementary frameworks and summarized in several reviews \cite{Lebed2017,Esposito2017,Olsen2018,Brambilla2020,Ali2017,Hosaka2016,Liu2019,ChenReview2023,Meng2023}. The picture was soon extended to the strange sector, where the $P_{cs}(4459)$ \cite{Aaij2021} and $P_{cs}(4338)$ \cite{Aaij2023} are well described as $\bar{D}^{*}\Xi_c$ molecules with hidden charm and a unit of strangeness~\cite{Wang2020}, with a further candidate reported in the $B_s^0\to J/\psi p\bar{p}$ channel \cite{LHCbPc4337}.

According to the discoveries of several pentaquark states, one might ask a natural question about how far the molecular spectrum implied by these symmetries extends in flavor space. Hadrons whose valence content goes beyond $q\bar{q}$ or $qqq$ are exotic, and among them the configurations built from four or five different flavors are the most distinctive. Since the discovery of the $X(3872)$ \cite{Choi2003}, a large family of such states has been established, including the doubly charmed tetraquark $T_{cc}^+$ found just below the $D^{*+}D^0$ threshold \cite{LHCbTcc1,LHCbTcc2}, which reinforces the relevance of near-threshold molecular dynamics across the heavy-flavor sector. The open-flavor tetraquarks $X_0(2900)$ and $X_1(2900)$ \cite{Aaij2020,LHCbX2900b} carry four different flavors and gave clear evidence for this kind of state. The next step is a hadron containing the five long-lived quark flavors, namely bottom, charm, strange, up and down. A pentaquark with quark content $udsc\bar{b}$ realizes this configuration, with open charm $C=+1$ and open bottomness carried by the anti-bottom quark, and it provides a clear flavor indicator for an experimental search.

Several groups have already pointed to such states. Ref.~\cite{Shen2022} studied the $B^{*}\Xi_c^{\prime}$ system through $t$-channel vector meson exchange and found bound states in the pseudoscalar-baryon and vector-baryon sectors treated separately. While Ref.~\cite{Peng2022} used $SU(3)$-flavor and heavy-flavor symmetry within a contact-range effective field theory to map the full multiplet structure of the LHCb pentaquarks onto their light- and heavy-flavor partners, the five-flavor states included. In addition, Ref.~\cite{Yu2019} analyzed the related $\Sigma_{bc}$ meson-baryon systems in the local hidden gauge framework. More recently,  Ref.~\cite{WangLiu2026} carried out a one-boson-exchange study of the $\bar{D}^{*}\Xi_b^{\prime,*}$ and $B^{*}\Xi_c^{\prime,*}$ systems with $S$-$D$ wave mixing and coupled channels. Furthermore, a molecular open-charm and open-bottom analysis was also given in Ref.~\cite{Lin2024}.

In this study, we adopt an approach in which the interaction kernel is constrained mainly by the relevant symmetries of the five-flavor pentaquark system. We start from the chiral unitary description given in Ref.~\cite{Xiao2019}, which reproduces the hidden-charm strange $P_{cs}$ states, and we organize the five-flavor interaction through three embedded symmetries. The chiral local hidden gauge (LHG) fixes the overall meson-baryon coupling through the Weinberg-Tomozawa (WT) term, with no free coupling. The Heavy-quark spin symmetry (HQSS) then relates the pseudoscalar-baryon and vector-baryon channels and links the whole tower of $J=1/2,3/2,5/2$ states. Heavy-quark flavor symmetry (HQFS) finally connects the five-flavor system to the charm sector that experiment has begun to map, by replacing the anti-charm quark in the meson with an anti-bottom quark while keeping the charm quark in the baryon. Because the recoupling that organizes the channels is blind to the heavy-quark flavor, the coefficient matrices carry over from charm to bottom unchanged, and the entire ten-state spectrum follows with a single adjustable parameter, the loop subtraction constant. This is the main advantage of this work. Although previous studies of the same system treat the pseudoscalar-baryon and vector-baryon sectors as separate scenarios~\cite{Shen2022}, our spectrum is combined together by the heavy-quark symmetries. 
Consequently, the relative pole positions of the states, their spin degeneracies, and the additional $\Xi_c^{*}$ partners emerge as predictions rather than inputs.

\section{Formalism}

\subsection{Chiral unitary approach}

We describe the $udsc\bar{b}$ pentaquarks as $S$-wave meson-baryon molecules generated dynamically in coupled channels. We work at fixed $J^P=\tfrac12^-$, $\tfrac32^-$, and $\tfrac52^-$, isospin $I=0$, open charm $C=+1$ and open bottom $B=+1$. The channels are read
\begin{align}
J=\tfrac{1}{2}\,:\ & B_c\Lambda,\ B_c^{*}\Lambda,\ B\Xic,\ B_s\Lamc,\ B\Xicp, \nonumber\\
 & B^{*}\Xic,\ B_s^{*}\Lamc,\ B^{*}\Xicp,\ B^{*}\Xicst, \nonumber\\
J=\tfrac{3}{2}\,:\ & B_c^{*}\Lambda,\ B^{*}\Xic,\ B_s^{*}\Lamc,\ B^{*}\Xicp,\ B\Xicst,\ B^{*}\Xicst, \nonumber\\
J=\tfrac{5}{2}\,:\ & B^{*}\Xicst .
\end{align}
The scattering amplitude follows from the coupled-channel Bethe-Salpeter equation in the on-shell factorization form \cite{Oset1998,Oller2001},
\begin{equation}
T(\sqs)=\big[1-V(\sqs)\,G(\sqs)\big]^{-1}V(\sqs),
\label{eq:bse}
\end{equation}
with $V$ the tree-level kernel and $G=\mathrm{diag}(G_1,\dots,G_n)$ the diagonal meson-baryon loop function. Bound and quasi-bound states show up as poles of $T$. The loop function of channel $l$ is regularized in dimensional regularization \cite{Oller2001},
\begin{equation}
\begin{split}
G_l(s)=\frac{2M_l}{16\pi^2}\bigg\{&a(\mu)+\ln\frac{M_l^2}{\mu^2}
+\frac{m_l^2-M_l^2+s}{2s}\ln\frac{m_l^2}{M_l^2}\\
&+\frac{q_l}{\sqs}\Big[\ln\!\big(s-(M_l^2-m_l^2)+2q_l\sqs\big)\\
&\quad+\ln\!\big(s+(M_l^2-m_l^2)+2q_l\sqs\big)\\
&\quad-\ln\!\big(-s+(M_l^2-m_l^2)+2q_l\sqs\big)\\
&\quad-\ln\!\big(-s-(M_l^2-m_l^2)+2q_l\sqs\big)\Big]\bigg\},
\end{split}
\label{eq:loop}
\end{equation}
where $m_l$ and $M_l$ are the meson and baryon masses and $q_l=\lambda^{1/2}(s,m_l^2,M_l^2)/(2\sqs)$ is the center-of-mass momentum, with $\lambda$ the Kallen function. The factor $2M_l$ comes from the baryon spinor normalization and matches the normalization chosen for $V$. We use a single subtraction constant $a(\mu)$ for all channels at $\mu=1\,$GeV.

A quasi-bound state is a pole of $T$ on the second Riemann sheet, reached by continuing the loop function of the open channels,
\begin{equation}
G_l^{\mathrm{II}}(s)=G_l^{\mathrm{I}}(s)+i\,\frac{2M_l\,q_l}{4\pi\sqs}\,,
\label{eq:secondsheet}
\end{equation}
while the closed channels are kept on the first sheet. The pole position $\sqs_R=M+i\Gamma/2$ is a zero of $\det(1-V G^{\mathrm{II}})$, and we report the half width as $\Gamma/2=|\mathrm{Im}\,\sqs_R|$. Near a pole, the amplitude factorizes as $T_{ij}(\sqs)\simeq g_i g_j/(\sqs-\sqs_R)$, which defines the complex couplings $g_i$ from the residue. We fix the global phase by the dominated channel, taking $g_{i_0}=\sqrt{R_{i_0 i_0}}$ and $g_i=R_{i i_0}/g_{i_0}$ with $R_{ij}$ the residue matrix.

The residues measure how molecular each pole is. The compositeness of channel $i$, i.e. the weight of that meson-baryon component in the state, is given by \cite{Weinberg1965,Gamermann2010,AcetiOset2012,HyodoJidoHosaka2012,Hyodo2013,HyodoPRL2013,Sekihara2015}
\begin{equation}
X_i=-\,g_i^2\,\frac{\partial G_i}{\partial\sqs}\bigg|_{\sqs_R},\qquad \sum_i X_i=1,
\label{eq:Xi}
\end{equation}
evaluated on the same Riemann sheet used for the pole. The sum rule holds for a state dynamically generated by an energy-independent kernel; the weak residual energy dependence of the WT term in Eq.~\eqref{eq:potential}, which the large bottom-meson mass renders almost constant, produces only a few-percent deviation. For a quasi-bound state $X_i$ is in general complex and is read as a weight rather than a strict probability \cite{Aceti2014}; for the narrow poles found here its imaginary part is negligible and $\mathrm{Re}\,X_i$ gives the molecular probability of the channel $i$ directly.

\subsection{Interaction from heavy-quark spin symmetry and the local hidden gauge}
\label{sec:HQSS-LHG}

The pentaquarks arise from the $S$-wave meson-baryon interaction, which has the WT form
\begin{equation}
V_{ij}(\sqs)=-\,\frac{p_i^0+p_j^0}{4f^2}\,C^{(J)}_{ij},\qquad p_i^0=\frac{s+m_i^2-M_i^2}{2\sqs},
\label{eq:potential}
\end{equation}
with $f$ the pion decay constant and $p_i^0$ the on-shell meson energy in channel $i$. The energy prefactor is the universal light meson-baryon strength, and all of the channel dynamics is carried by the dimensionless coefficient matrix $C^{(J)}_{ab}$. Two symmetries determine this matrix without adjustable couplings. HQSS specifies the coupled channels and their relative weights, whereas the LHG fixes the limited independent strengths. We address these two factors in the latter.

HQSS is the statement that, in the heavy-quark limit, the strong interaction is invariant under independent spin rotations of the heavy quark and the heavy antiquark. The spin of the heavy pair $S_{Q\bar{Q}}$ and the total angular momentum $L$ of the light degrees of freedom are then separately conserved, and in the basis labeled by $(L,S_{Q\bar{Q}})$ the interaction is diagonal and independent of both $S_{Q\bar{Q}}$ and the total spin $J$. Only two reduced matrix elements survive, a $4\times4$ symmetric matrix for the $L=\tfrac12$ light configurations and a single number for the $L=\tfrac32$ configuration. We label the $L=\tfrac12$ configurations by $1$ for the $B_c^{*}\Lambda$ content, $2$ for $B^{*}\Xi_c$, $3$ for $B_s^{*}\Lambda_c$ and $4$ for $B^{*}\Xi_c'$, and the $L=\tfrac32$ configuration is the $B^{*}\Xi_c^{*}$ content. The physical meson-baryon channels are projected onto this HQSS basis by a Racah recoupling, written as a $9j$ symbol \cite{Xiao2013,Xiao2019},
\begin{equation}
\begin{split}
R^{(J)}_a(L,S_{Q\bar{Q}})=\ &\Big[(2S_{Q\bar{Q}}+1)(2L+1)(2j^a_M+1)(2j^a_B+1)\Big]^{1/2}\\
&\times\begin{Bmatrix}\ell^a_M & \ell^a_B & L\\ \tfrac12 & \tfrac12 & S_{Q\bar{Q}}\\ j^a_M & j^a_B & J\end{Bmatrix},
\end{split}
\label{eq:9j}
\end{equation}
where $\ell_M$ and $\ell_B$ are the light angular momenta in the meson and the baryon, $j_M$ and $j_B$ are the meson and baryon spins, and the heavy-quark spins are $\tfrac12$. The $B_c$ and $B_c^{*}$ mesons carry the whole heavy quark-antiquark pair, so they are exact HQSS eigenstates with $\ell_M=0$, $L=\tfrac12$ and $S_{Q\bar{Q}}=0$ or $1$, for which $R^{(J)}_a=\delta_{L,1/2}\,\delta_{S_{Q\bar{Q}},S^a_{Q\bar{Q}}}$.

The LHG fixes the size of the two reduced elements through vector-meson exchange between the meson and the baryon\textcolor{blue}{~\cite{Bando1985,Bando1988,Meissner1988}}. The dominated term is the exchange of a light vector $\rho$, $\omega$ or $\phi$, for which the heavy antiquark is a spectator. It carries the leading WT strength and is automatically independent of the heavy-quark spin, in line with the symmetry above. A transition that changes the heavy content of the meson instead needs the exchange of a heavy vector $B^{*}$, whose large mass suppresses it by the factor $\gamma$. Writing the reduced matrix as $\hat{\mu}_{ij}=-F\,\kappa^{(L)}_{ij}$ with $F=(p_i^0+p_j^0)/(4f^2)$, the light-vector exchange gives 
\begin{equation}
\kappa^{(1/2)}=
\begin{pmatrix}
0 & \sqrt{\tfrac{2}{3}}\,\gamma & -\tfrac{2}{\sqrt3}\,\gamma & \sqrt{2}\,\gamma\\[2pt]
\sqrt{\tfrac{2}{3}}\,\gamma & 1 & \sqrt2 & 0\\[2pt]
-\tfrac{2}{\sqrt3}\,\gamma & \sqrt2 & 0 & 0\\[2pt]
\sqrt2\,\gamma & 0 & 0 & 1
\end{pmatrix},
\qquad
\gamma\equiv\frac{m_V^2}{m_{B^{*}}^2}\,,
\label{eq:kappa}
\end{equation}
and $\kappa^{(3/2)}=1$.

The unsuppressed units and the entries $\sqrt2$ are in the diagonal $L=\tfrac12$ block that couples $B^{*}\Xi_c$ to $B_s^{*}\Lambda_c$, while every element that connects to the $B_c^{*}\Lambda$ content carries a factor $\gamma$ and is therefore small. Combining the recoupling and the reduced elements gives the full coefficient matrix,
\begin{equation}
C^{(J)}_{ij}=\sum_{L,S_{Q\bar{Q}}}R^{(J)}_i(L,S_{Q\bar{Q}})\,\kappa^{(L)}_{c(i)c(j)}\,R^{(J)}_j(L,S_{Q\bar{Q}}),
\label{eq:Cmatrix}
\end{equation}
where $c(i)\in\{1,2,3,4\}$ is the configuration index of channel $i$. The three coefficient matrices are given by $J= \tfrac12,\,\tfrac32$, and $\tfrac52$, respectively, as.

For $J=\tfrac12$ sector, the channels ($B_c\Lambda$, $B_c^{*}\Lambda$, $B\Xic$, $B_s\Lamc$, $B\Xicp$, $B^{*}\Xic$, $B_s^{*}\Lamc$, $B^{*}\Xicp$, $B^{*}\Xicst$) lead to the $9\times9$ matrix
\begin{widetext}
\begin{equation}
C^{(1/2)}=
\begin{pmatrix}
0 & 0 & -\tfrac{1}{\sqrt6}\gamma & \tfrac{1}{\sqrt3}\gamma & \tfrac{1}{\sqrt2}\gamma & \tfrac{1}{\sqrt2}\gamma & -\gamma & \tfrac{1}{\sqrt6}\gamma & \tfrac{2}{\sqrt3}\gamma\\[3pt]
0 & 0 & \tfrac{1}{\sqrt2}\gamma & -\gamma & \tfrac{1}{\sqrt6}\gamma & \tfrac{1}{\sqrt6}\gamma & -\tfrac{1}{\sqrt3}\gamma & \tfrac{5}{3\sqrt2}\gamma & -\tfrac{2}{3}\gamma\\[3pt]
-\tfrac{1}{\sqrt6}\gamma & \tfrac{1}{\sqrt2}\gamma & 1 & \sqrt2 & 0 & 0 & 0 & 0 & 0\\[3pt]
\tfrac{1}{\sqrt3}\gamma & -\gamma & \sqrt2 & 0 & 0 & 0 & 0 & 0 & 0\\[3pt]
\tfrac{1}{\sqrt2}\gamma & \tfrac{1}{\sqrt6}\gamma & 0 & 0 & 1 & 0 & 0 & 0 & 0\\[3pt]
\tfrac{1}{\sqrt2}\gamma & \tfrac{1}{\sqrt6}\gamma & 0 & 0 & 0 & 1 & \sqrt2 & 0 & 0\\[3pt]
-\gamma & -\tfrac{1}{\sqrt3}\gamma & 0 & 0 & 0 & \sqrt2 & 0 & 0 & 0\\[3pt]
\tfrac{1}{\sqrt6}\gamma & \tfrac{5}{3\sqrt2}\gamma & 0 & 0 & 0 & 0 & 0 & 1 & 0\\[3pt]
\tfrac{2}{\sqrt3}\gamma & -\tfrac{2}{3}\gamma & 0 & 0 & 0 & 0 & 0 & 0 & 1
\end{pmatrix}.
\label{eq:C12}
\end{equation}
\end{widetext}
For $J=\tfrac32$, the channels $(B_c^{*}\Lambda$, $B^{*}\Xic$, $B_s^{*}\Lamc$, $B^{*}\Xicp$, $B\Xicst$, $B^{*}\Xicst)$, we obtain
\begin{equation}
C^{(3/2)}=
\begin{pmatrix}
0 & \sqrt{\tfrac{2}{3}}\,\gamma & -\tfrac{2}{\sqrt3}\,\gamma & -\tfrac{\sqrt2}{3}\,\gamma & \sqrt{\tfrac{2}{3}}\,\gamma & \tfrac{\sqrt{10}}{3}\,\gamma\\[3pt]
\sqrt{\tfrac{2}{3}}\,\gamma & 1 & \sqrt2 & 0 & 0 & 0\\[3pt]
-\tfrac{2}{\sqrt3}\,\gamma & \sqrt2 & 0 & 0 & 0 & 0\\[3pt]
-\tfrac{\sqrt2}{3}\,\gamma & 0 & 0 & 1 & 0 & 0\\[3pt]
\sqrt{\tfrac{2}{3}}\,\gamma & 0 & 0 & 0 & 1 & 0\\[3pt]
\tfrac{\sqrt{10}}{3}\,\gamma & 0 & 0 & 0 & 0 & 1
\end{pmatrix}.
\label{eq:C32}
\end{equation}
The single $J=\tfrac52$ channel is $B^{*}\Xi_c^{*}$, and the coefficient matrix reduces to a single number as,
\begin{equation}
C^{(5/2)}=1 .
\label{eq:C52}
\end{equation}
The $C^{(1/2)}$ and $C^{(3/2)}$ matrices of the leading WT interaction reveal that the $B^{*}\Xi_c$ and $B_s^{*}\Lambda_c$ channels are coupled, whereas the $B_c^{*}\Lambda$ channels enter only through the $\gamma$-suppressed elements and completely decouple in the perfect heavy-quark limit $\gamma\to0$. For the $J=\tfrac52$ channel, $C^{(5/2)}$ carries the same unit diagonal strength. This matrix structure organizes the spectrum shown in the latter.

\subsection{Heavy-quark flavor symmetry and the five-flavor system}

The construction above describes the hidden-charm strange $P_{cs}$ states once the meson masses are taken in the charm sector and the suppression factor is $\gamma_c=m_V^2/m_{D^{*}}^2$ \cite{Xiao2019,Xiao2013}. Heavy-quark flavor symmetry lets us carry the same dynamics to the five-flavor sector. At leading order in the heavy-quark expansion a heavy quark enters the interaction only as a static color source \cite{Isgur1989,Isgur1990,Neubert1994,ManoharWise2000}, through the effective Lagrangian\textcolor{blue}{~\cite{Wise1992,Yan1992,Casalbuoni1997}}
\begin{equation}
\mathcal{L}_{Q}=\bar{h}^{(Q)}_v\,(iv\!\cdot\!D)\,h^{(Q)}_v+\mathcal{O}(1/m_Q),
\label{eq:hqet}
\end{equation}
which has the same form for $Q=c$ and $Q=b$. The light degrees of freedom therefore cannot tell the two heavy flavors apart, and the reduced matrix elements of the interaction are flavor blind,
\begin{equation}
\kappa^{(L)}_{ij}\big|_{\bar{b}}=\kappa^{(L)}_{ij}\big|_{\bar{c}}\equiv\kappa^{(L)}_{ij},
\qquad
C^{(J)}_{ij}\big|_{\bar{b}}=C^{(J)}_{ij}\big|_{\bar{c}} .
\label{eq:univ}
\end{equation}
We exploit this by replacing the anti-charm quark in the meson with an anti-bottom quark while keeping the charm quark in the baryon. The heavy-meson multiplets can be mapped as
\begin{equation}
(\eta_c,\,J/\psi,\,\bar{D},\,\bar{D}^{*},\,\bar{D}_s,\,\bar{D}_s^{*})\ \longrightarrow\ (B_c,\,B_c^{*},\,B,\,B^{*},\,B_s,\,B_s^{*})\,,
\label{eq:map}
\end{equation}
while the baryons $\Lambda$, $\Lambda_c$, $\Xi_c$, $\Xi_c'$ and $\Xi_c^{*}$ keep their charm quark unchanged. These pairs of mesons and baryons produce the five-flavor $udsc\bar{b}$ system.

By Eq.~\eqref{eq:univ} the recoupling of Eq.~\eqref{eq:9j} and the coefficient matrices of Eqs.~\eqref{eq:C32} and \eqref{eq:C12} carry over unchanged, so the entire flavor dependence of the kernel is concentrated in the hadron masses and in the single ratio that governs the heavy-vector exchange,
\begin{equation}
\gamma_Q=\frac{m_V^2}{m_{P_Q^{*}}^2},\qquad
\frac{\gamma_c}{\gamma_b}=\frac{m_{B^{*}}^2}{m_{D^{*}}^2}\simeq7,
\label{eq:gamma}
\end{equation}
with $P_Q^{*}$ the heavy vector meson, $D^{*}$ for charm and $B^{*}$ for bottom. The heavy-content-changing transitions, already a small effect in charm, are therefore suppressed about seven times more strongly here. The final interaction kernel $V^{(b)}_{ij}=-(p_a^0+p_b^0)\,C^{(J)}_{ij}(\gamma_b)/(4f^2)$ is in this way fixed by the charm one without new coupling, which makes the prediction a symmetry-constrained extrapolation of the dynamics that already reproduces the observed charm states rather than an independent model.

In practice, only the input changes, that is, the meson masses shift to those of the bottom sector, the suppression factor decreases from $\gamma_c\simeq0.16$ to $\gamma_b\simeq0.023$, and the subtraction constant is adjusted. The naive charm value $a=-2.09$ is not bound here, because $m_{B^{*}}>M_{\Xi_c^{*}}$ reverses the role of the meson and the baryon in the threshold loop. More concretely, the loop function of Eq.~\eqref{eq:loop} carries the baryon normalization factor $2M_l$ and the logarithm $\ln(M_l^2/\mu^2)$, so at fixed $a(\mu)$ the size of $G$ is controlled by the baryon mass, namely, when the meson becomes heavier than the baryon. As for the bottom mesons here, a more negative $a$ is required to reach the same loop strength and hence the same binding. We adopt $a(\mu=1\,\mathrm{GeV})=-3.1$, the value used by Ref.~\cite{Shen2022} for this exactly the same system, which is also intermediate between the hidden-charm value $-2.09$ \cite{Xiao2019b} and the hidden-bottom value $-3.71$ \cite{Wu2012} and keeps the binding moderate. This is the single scheme parameter of the calculation, and the absolute masses scale with it.
For this reason, the most generic predictions are the existence of the pole pattern, the HQSS degeneracies, and the dominated channel assignments. The absolute binding energies should be read together with the subtraction-constant dependence shown below.

The numerical inputs are as follows, with all masses in MeV. The coupling constant is set by the pion decay constant $f_\pi=93$ and the averaged light-vector mass $m_V=800$, the heavy-vector suppression factor is $\gamma=m_V^2/m_{B^{*}}^2=0.023$, and the loop subtraction constant is $a(\mu)=-3.1$ at $\mu=1\,$GeV. The bottom-meson masses are $m_{B_c}=6274.5$ and $m_{B_c^{*}}=6331$, the latter a theory average since the state is not yet well established, together with $m_B=5279.4$, $m_{B_s}=5366.9$, $m_{B^{*}}=5324.7$ and $m_{B_s^{*}}=5415.4$. The baryon masses are $M_\Lambda=1115.68$ and $M_{\Lambda_c}=2286.46$ for the $\Lambda$ states, and $M_{\Xi_c}=2469.42$, $M_{\Xi_c'}=2578.8$ and $M_{\Xi_c^{*}}=2645.9$ for the charmed cascades.

\section{Results}

We solve Eq.~\eqref{eq:bse} in each fixed-$J$ coupled-channel space, search for $\det[1-VG^{\mathrm{II}}]=0$ on the appropriate unphysical Riemann sheets, and extract the pole positions, half widths, and residues. We use the convention $\sqrt{s_p}=M+i\Gamma/2$ for a resonance pole, as in~\cite{Xiao2019}. 
To verify our numerical framework, we have reproduced the hidden-charm strange spectrum of~\cite{Xiao2019} with charm inputs, recovering the poles in numerical agreement with the published results. Using the same manner, numerical calculation, this yields the five-flavor spectrum with the bottom mesons input. 

\begin{figure}[t]
\centering
\includegraphics[width=0.46\textwidth]{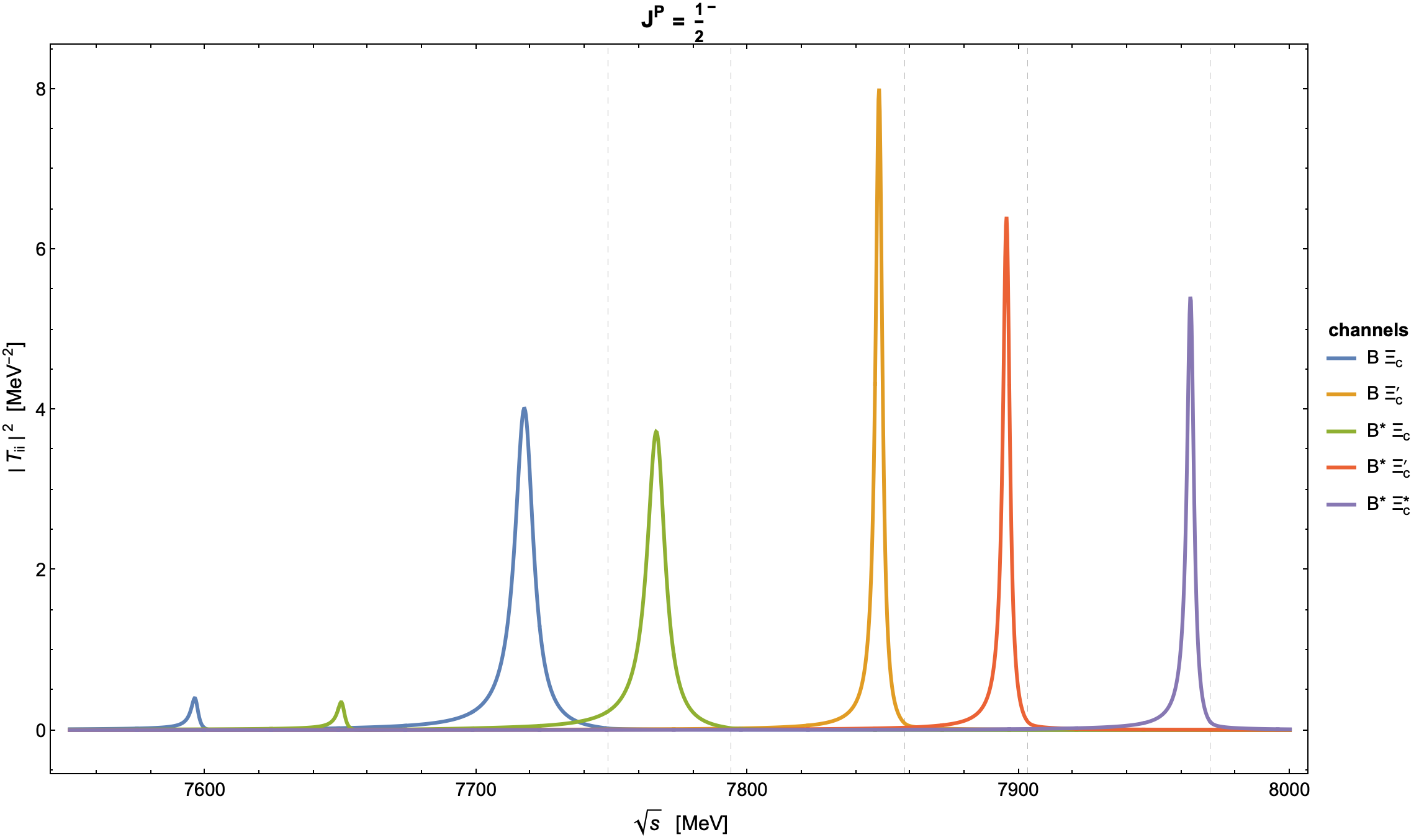}
\caption{Squared moduli of the diagonal amplitudes $|T_{ii}|^2$ for $J^P=1/2^-$ as a function of $\sqs$. The $B\Xi_c$ and $B^{*}\Xi_c$ curves each display two peaks, the higher one being the dominated molecule of Table~\ref{tab:spectrum} and the lower one the deeply bound $B_s^{(*)}\Lambda_c$ partner discussed in the text.}
\label{fig:J12}
\end{figure}

\begin{figure}[t]
\centering
\includegraphics[width=0.46\textwidth]{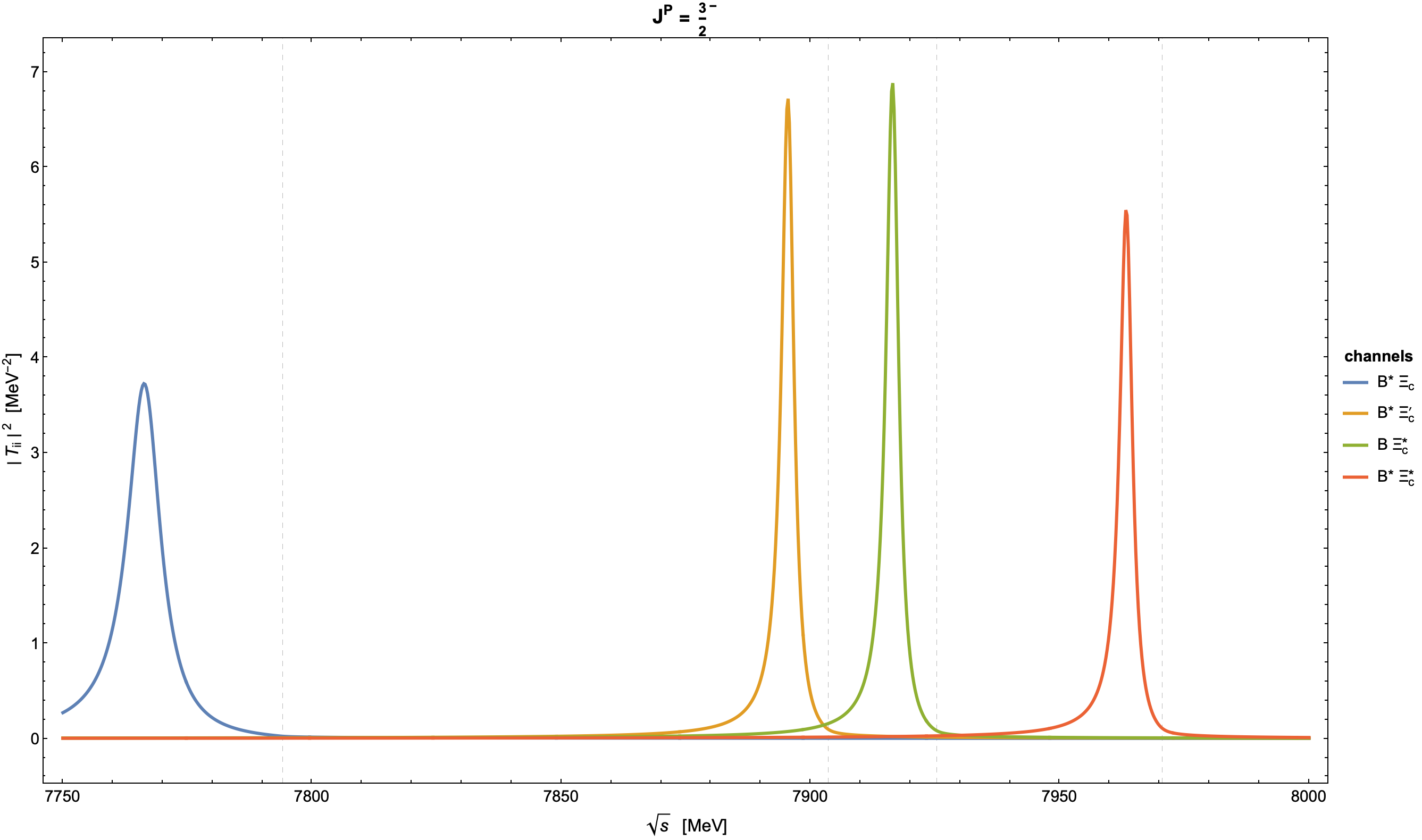}
\caption{Squared moduli of the diagonal amplitudes $|T_{ii}|^2$ for $J^P=3/2^-$ as a function of $\sqs$.}
\label{fig:J32}
\end{figure}

For the $I=0$ sector, ten threshold-associated poles emerge with the same heavy-quark spin multiplet pattern as in charm. Collecting all heavy-quark spin degeneracies of the resonances, these ten $(J^P,\text{state})$ entries correspond to six distinct mass multiplets. Their masses and half widths are given in Table~\ref{tab:spectrum}. All of them are narrow, with $\Gamma/2$ below about $3$ MeV. The small widths reflect a combination of phase space ($\Gamma_i \propto q_i\,|g_i|^2$) each open-channel partial width, the weak coupling to the lower $B_c^{*}\Lambda$ channels, and, for the two deepest upper poles, the finite but small coupling to the lower $B_s^{*}\Lambda_c$ channels. The amplitudes $|T_{ii}|^2$ are shown in Fig.~\ref{fig:J12} for $J=1/2$ and in Fig.~\ref{fig:J32} for $J=3/2$, respectively.

\begin{table}[t]
\centering
\caption{Predicted $udsc\bar{b}$ molecular spectrum and compositeness for $a(\mu=1\,\mathrm{GeV})=-3.1$. $M$ is the mass and $\Gamma/2$ the half width (both in MeV), with the dominant channel and the relevant threshold listed. $X_{\rm dom}=\mathrm{Re}\,X_i$ is the compositeness of the dominant channel and $X_{B_s^{(*)}\Lambda_c}$ the weight of the lower open partner channel; the total $\sum_i X_i$ is unity to within a percent in all cases.}
\label{tab:spectrum}
{\footnotesize\setlength{\tabcolsep}{3pt}
\begin{tabular}{ccccccc}
\toprule
$J^P$ & dominant channel & threshold & $M$ & $\Gamma/2$ & \textcolor{black}{$X_{\rm dom}$} & \textcolor{black}{$X_{B_s^{(*)}\Lambda_c}$}\\
\midrule
$1/2^-$ & $B\Xi_c$        & 7748.8 & 7717.8 & 2.6 & \textcolor{black}{0.98} & \textcolor{black}{0.02}\\
$1/2^-$ & $B^{*}\Xi_c$    & 7794.1 & 7766.5 & 2.5 & \textcolor{black}{0.98} & \textcolor{black}{0.02}\\
$1/2^-$ & $B\Xi_c'$       & 7858.2 & 7848.5 & $\lesssim1$ & \textcolor{black}{0.99} & \textcolor{black}{$<0.01$}\\
$1/2^-$ & $B^{*}\Xi_c'$   & 7903.5 & 7895.5 & $\lesssim1$ & \textcolor{black}{0.99} & \textcolor{black}{$<0.01$}\\
$1/2^-$ & $B^{*}\Xi_c^{*}$& 7970.6 & 7963.4 & $\lesssim1$ & \textcolor{black}{0.99} & \textcolor{black}{$<0.01$}\\
$3/2^-$ & $B^{*}\Xi_c$    & 7794.1 & 7766.5 & 2.5 & \textcolor{black}{0.98} & \textcolor{black}{0.02}\\
$3/2^-$ & $B^{*}\Xi_c'$   & 7903.5 & 7895.5 & $\lesssim1$ & \textcolor{black}{0.99} & \textcolor{black}{$<0.01$}\\
$3/2^-$ & $B\Xi_c^{*}$    & 7925.3 & 7916.5 & $\lesssim1$ & \textcolor{black}{0.99} & \textcolor{black}{$<0.01$}\\
$3/2^-$ & $B^{*}\Xi_c^{*}$& 7970.6 & 7963.4 & $\lesssim1$ & \textcolor{black}{0.99} & \textcolor{black}{$<0.01$}\\
$5/2^-$ & $B^{*}\Xi_c^{*}$& 7970.6 & 7963.4 & 0 & \textcolor{black}{0.99} & \textcolor{black}{0}\\
\bottomrule
\end{tabular}}
\end{table}

The binding pattern follows the eigenvalue structure of $C^{(J)}$. The deeply bound $B\Xi_c$ and $B^{*}\Xi_c$ states emerge from the attractive $-2F$ factor of the coupled $B^{*}\Xi_c$ to $B_s^{*}\Lambda_c$ channels, while the $B\Xi_c'$, $B^{*}\Xi_c'$, $B\Xi_c^{*}$ and $B^{*}\Xi_c^{*}$ states bind through the weaker diagonal $-F$ and sit only a few MeV below their thresholds.

The full set of complex couplings is given in Table~\ref{tab:coup12} for $J=1/2$ and in Table~\ref{tab:coup32} for $J=3/2$, in the same structure as the hidden-charm study of Ref.~\cite{Xiao2019}. Each state is strongly dominated by a single meson-baryon channel, which fixes its molecular nature, and the pattern of dominant channels is identical in the two spin sectors wherever a channel is shared. The lightest pole $J=1/2$ is almost entirely bound to $B\Xi_c$, with $|g_{B\Xi_c}|=2.87$ and all other moduli below $0.5$, so it is a $B\Xi_c$ molecule bound by about $31$ MeV. The pole at $7766.5$ MeV has $|g_{B^{*}\Xi_c}|=2.79$ and is the $B^{*}\Xi_c$ molecule, bound by about $28$ MeV, and it appears with the same mass and the same coupling in both $J=1/2$ and $J=3/2$, the first member of a heavy-quark spin doublet. The small but nonzero couplings of the two deeply bound states to the open $B_s^{*}\Lambda_c$ channels, $|g_{B_s^{*}\Lambda_c}|=0.48$, are what give them their finite widths, while the four weakly bound $-F$ states couple to the closed $B_c^{*}\Lambda$ channels only through the $\gamma$-suppressed elements and are correspondingly narrow.

\begin{table*}[t]
\centering
\caption{Dimensionless complex couplings $g_i$ and their moduli $|g_i|$ for the five $(I=0,J^P=1/2^-)$ poles of the $udsc\bar{b}$ system at $a(\mu=1\,\mathrm{GeV})=-3.1$. Each pole is labeled by its second Riemann sheet position $\sqrt{s_p}=M+i\Gamma/2$, in MeV, with positive imaginary part following the convention of Ref.~\cite{Xiao2019}. The dominated coupling of each state is set in bold.}
\label{tab:coup12}
\footnotesize
\setlength{\tabcolsep}{3.2pt}
\begin{tabular}{l ccccccccc}
\toprule
 & $B_c\Lambda$ & $B_c^{*}\Lambda$ & $B\Xi_c$ & $B_s\Lambda_c$ & $B\Xi_c'$ & $B^{*}\Xi_c$ & $B_s^{*}\Lambda_c$ & $B^{*}\Xi_c'$ & $B^{*}\Xi_c^{*}$\\
\midrule
\multicolumn{10}{l}{$\sqrt{s_p}=7717.8+i\,2.61$}\\
\ $g_i$ & $0.02-i0.01$ & $-0.04+i0.01$ & $\bm{2.87-i0.07}$ & $0.30+i0.38$ & $0.00$ & $0.00$ & $0.00$ & $0.00$ & $0.00$\\
\ $|g_i|$ & $0.03$ & $0.04$ & $\bm{2.87}$ & $0.48$ & $0.00$ & $0.00$ & $0.00$ & $0.00$ & $0.00$\\[3pt]
\multicolumn{10}{l}{$\sqrt{s_p}=7766.5+i\,2.53$}\\
\ $g_i$ & $-0.04+i0.01$ & $-0.02+i0.01$ & $0.00$ & $0.00$ & $0.00$ & $\bm{2.79-i0.07}$ & $0.29+i0.38$ & $0.00$ & $0.00$\\
\ $|g_i|$ & $0.04$ & $0.02$ & $0.00$ & $0.00$ & $0.00$ & $\bm{2.79}$ & $0.48$ & $0.00$ & $0.00$\\[3pt]
\multicolumn{10}{l}{$\sqrt{s_p}=7848.5+i\,0.02$}\\
\ $g_i$ & $0.04$ & $0.02$ & $0.00$ & $0.00$ & $\bm{2.07}$ & $0.00$ & $0.00$ & $0.00$ & $0.00$\\
\ $|g_i|$ & $0.04$ & $0.02$ & $0.00$ & $0.00$ & $\bm{2.07}$ & $0.00$ & $0.00$ & $0.00$ & $0.00$\\[3pt]
\multicolumn{10}{l}{$\sqrt{s_p}=7895.5+i\,0.04$}\\
\ $g_i$ & $0.02$ & $0.06$ & $0.00$ & $0.00$ & $0.00$ & $0.00$ & $0.00$ & $\bm{1.97}$ & $0.00$\\
\ $|g_i|$ & $0.02$ & $0.06$ & $0.00$ & $0.00$ & $0.00$ & $0.00$ & $0.00$ & $\bm{1.97}$ & $0.00$\\[3pt]
\multicolumn{10}{l}{$\sqrt{s_p}=7963.4+i\,0.05$}\\
\ $g_i$ & $0.05$ & $-0.03$ & $0.00$ & $0.00$ & $0.00$ & $0.00$ & $0.00$ & $0.00$ & $\bm{1.89}$\\
\ $|g_i|$ & $0.05$ & $0.03$ & $0.00$ & $0.00$ & $0.00$ & $0.00$ & $0.00$ & $0.00$ & $\bm{1.89}$\\[3pt]
\bottomrule
\end{tabular}
\end{table*}

\begin{table*}[t]
\centering
\caption{Same as Table~\ref{tab:coup12} for the four $(I=0,J^P=3/2^-)$ poles.}
\label{tab:coup32}
\footnotesize
\setlength{\tabcolsep}{5pt}
\begin{tabular}{l cccccc}
\toprule
 & $B_c^{*}\Lambda$ & $B^{*}\Xi_c$ & $B_s^{*}\Lambda_c$ & $B^{*}\Xi_c'$ & $B\Xi_c^{*}$ & $B^{*}\Xi_c^{*}$\\
\midrule
\multicolumn{7}{l}{$\sqrt{s_p}=7766.5+i\,2.53$}\\
\ $g_i$ & $-0.05+i0.01$ & $\bm{2.79-i0.07}$ & $0.29+i0.38$ & $0.00$ & $0.00$ & $0.00$\\
\ $|g_i|$ & $0.05$ & $\bm{2.79}$ & $0.48$ & $0.00$ & $0.00$ & $0.00$\\[3pt]
\multicolumn{7}{l}{$\sqrt{s_p}=7895.5+i\,0.01$}\\
\ $g_i$ & $-0.02$ & $0.00$ & $0.00$ & $\bm{1.97}$ & $0.00$ & $0.00$\\
\ $|g_i|$ & $0.02$ & $0.00$ & $0.00$ & $\bm{1.97}$ & $0.00$ & $0.00$\\[3pt]
\multicolumn{7}{l}{$\sqrt{s_p}=7916.5+i\,0.02$}\\
\ $g_i$ & $0.04$ & $0.00$ & $0.00$ & $0.00$ & $\bm{1.99}$ & $0.00$\\
\ $|g_i|$ & $0.04$ & $0.00$ & $0.00$ & $0.00$ & $\bm{1.99}$ & $0.00$\\[3pt]
\multicolumn{7}{l}{$\sqrt{s_p}2=7963.4+i\,0.03$}\\
\ $g_i$ & $0.05$ & $0.00$ & $0.00$ & $0.00$ & $0.00$ & $\bm{1.89}$\\
\ $|g_i|$ & $0.05$ & $0.00$ & $0.00$ & $0.00$ & $0.00$ & $\bm{1.89}$\\[3pt]
\bottomrule
\end{tabular}
\end{table*}

The molecular character of the poles is made quantitative by the compositeness of Eq.~\eqref{eq:Xi}, listed alongside the spectrum in Table~\ref{tab:spectrum}. Every threshold-associated state is saturated by a single meson-baryon channel, that is, the dominant weight is $X_{\rm dom}=0.98$ for the two deeply bound $B^{*}\Xi_c$ poles and $X_{\rm dom}\simeq0.99$ for the six weakly bound $-F$ states, with $\sum_i X_i$ within a percent of unity in all cases. The only sizeable secondary weight is the $1$--$2\%$ carried by the open $B_s^{*}\Lambda_c$ channels in the two $B^{*}\Xi_c$ states, the same coupling that gives them their finite widths. All $\gamma$-suppressed components are below $10^{-3}$. The compositeness therefore confirms that these are genuine $S$-wave molecules and not mixtures, and that the dominant-channel assignment of Table~\ref{tab:spectrum} is well defined, the imaginary part of $X_i$ being negligible for these narrow poles.

Table~\ref{tab:adep} lists how the masses move with the subtraction constant over the range $a=-2.9$ to $-3.2$, which brackets the values used by Ref.~\cite{Shen2022}. As $a$ decreases, the interaction becomes more attractive, and every state moves further below its threshold. Near $a=-2.9$ the weakly bound states sit essentially at their thresholds, so this value marks the onset of binding for the $-F$ states. The deeply bound $B\Xi_c$ and $B^{*}\Xi_c$ states survive throughout the range.

\begin{table}[t]
\centering
\caption{Pole masses (MeV) of the $udsc\bar{b}$ states as a function of the subtraction constant $a(\mu=1\,\mathrm{GeV})$.}
\label{tab:adep}
{\small\setlength{\tabcolsep}{4pt}
\begin{tabular}{lcccc}
\toprule
state ($J^P$) & $-2.9$ & $-3.0$ & $-3.1$ & $-3.2$\\
\midrule
$B\Xi_c\,(1/2^-)$ & 7747.5 & 7737.8 & 7717.8 & 7686.0\\
$B^{*}\Xi_c\,(1/2^-,3/2^-)$ & 7793.4 & 7785.1 & 7766.5 & 7736.3\\
$B\Xi_c'\,(1/2^-)$ & 7858.3 & 7857.4 & 7848.5 & 7828.9\\
$B^{*}\Xi_c'\,(1/2^-,3/2^-)$ & 7903.6 & 7903.1 & 7895.5 & 7877.3\\
$B\Xi_c^{*}\,(3/2^-)$ & 7926.3 & 7924.7 & 7916.5 & 7897.7\\
$B^{*}\Xi_c^{*}\,(1/2^-,3/2^-,5/2^-)$ & 7971.1 & 7970.4 & 7963.4 & 7945.9\\
\bottomrule
\end{tabular}}
\end{table}

\section{Physical implications and discussion}

\subsection{Comparison with previous work and the role of the symmetries}

Ref.~\cite{Shen2022} studied the same $udsc\bar{b}$ system with the same driving interaction, light vector exchange in the $t$-channel within the Bethe-Salpeter equation, the same scale $\mu=1\,$GeV and the same subtraction constant $a=-3.1$. The dynamical input is therefore essentially identical, and the comparison isolates the effect of the symmetry structure. Two differences matter. They solve the pseudoscalar-baryon and the vector-baryon channels as two separate four-channel scenarios, with no relation between them, whereas heavy-quark spin symmetry combines all of our channels into a single calculation. They also omit the spin-$\tfrac32$, $\Xi_c^{*}$ baryon, so they have neither a $J=5/2$ state nor any $B^{*}\Xi_c^{*}$ channel. In the formalism of Section~\ref{sec:HQSS-LHG}, the separate-sector treatment is the limit in which the recoupling between the elements $S_{Q\bar{Qj}}=0$ and $S_{Q\bar{Q}}=1$ is turned off by hand.

Both analyzes yield consistent results within their overlap domain. The four states common to both poles $B\Xi_c$ and $B\Xi_c'$, with poles $J^P=1/2^-$, as well as poles $B^{*}\Xi_c$ and $B^{*}\Xi_c'$, have masses that agree within roughly 2 MeV. 
Their couplings are slightly different by about 0.05, with \(|g_{B\Xi_c}| = 2.87\) compared to \(2.85\), and \(|g_{B^{*}\Xi_c}| = 2.79\) compared to \(2.77\). The slightly narrower widths in this work for the deeply bound states arise from the treatment of \(B_c^{*}\Lambda\) heavy-content-changing transitions, where we employ the constant factor \(\gamma=m_V^2/m_{B^{*}}^2\) instead of the full \(B^{*}\) propagator used in Ref.~\cite{Shen2022}.

This agreement is more than a numerical check. It shows that the separate-sector calculation is recovered as a limit of this work. The symmetries we consider here are controlled rather than ad hoc. The heavy-quark spin symmetry that links the two sectors does not cost a new parameter. It predicts the spin partners for both mesons and baryons. The smallness of the calculated cross-coupling explains the close agreement between the separate-sector masses. The main advantage of this approach is its predictive structure, namely, the prediction of ten rather than four states from a single subtraction constant. These are organized into heavy-quark spin multiplets with definite degeneracy relations and are connected by heavy-quark flavor symmetry to the charm states empirically supported by LHCb data.

\subsection{Heavy-quark spin multiplets}

A direct test of the heavy-quark spin symmetry is that the states group into multiplets that are degenerate up to the symmetry breaking induced by physical hadron masses and loop effects. We find three of them. The $B^{*}\Xi_c$ configuration gives a $J=1/2,3/2$ doublet at $7766.5$ MeV, with the same dominant coupling $|g_{B^{*}\Xi_c}|=2.79$ in both members, as Tables~\ref{tab:coup12} and \ref{tab:coup32} show. The $B^{*}\Xi_c'$ configuration gives a second doublet at $7895.5$ MeV. The $B^{*}\Xi_c^{*}$ configuration gives a $J=1/2,3/2,5/2$ triplet, which degenerates better than $1$ MeV at $7963.4$ MeV. The remaining states, the poles $B\Xi_c$ and $B\Xi_c'$ for $J=1/2$ and the pole $B\Xi_c^{*}$ for $J=3/2$, have no spin partner, because a pseudoscalar meson fixes the total spin once the baryon spin is chosen. The locations of the multiplets are not free parameters. They are inherited from the physical thresholds, in particular the $\Xi_c'$--$\Xi_c^{*}$ baryon splitting and the meson splitting $m_{B^{*}}-m_B$, which encode heavy-quark symmetry breaking of order $\Lambda_{\mathrm{QCD}}^2/m_Q$. This degeneracy pattern is a clear experimental signal of the heavy-quark symmetries, and it is absent by construction from any treatment that does not connect the pseudoscalar-baryon and vector-baryon channels.

The predictions of masses are also consistent with the broader symmetry survey of Ref.~\cite{Peng2022}, who place the $P_{cb}^{N}$ and related five-flavor states in the same $7.77$ to $7.91$ GeV window from $SU(3)$-flavor and heavy-flavor symmetry within a contact-range effective field theory. The one-boson-exchange study of Ref.~\cite{WangLiu2026} finds bound $B\Xi_c$ and $B^{*}\Xi_c$ states with the same quantum numbers, and through heavy-quark flavor symmetry they relate the $B\Xi_c$ states to the bottom partner of the observed $P_{cs}(4338)$. The qualitative spectrum, an octet-like pattern of narrow isoscalar states ordered by the meson and baryon content, is common to all of these approaches, which strengthens the case for the existence of five-flavor pentaquark states.

\subsection{Coupled-channel induced binding}

A feature absent in the charm sector appears in the bottom one. The $|T_{B\Xi_c}|^2$ and $|T_{B^{*}\Xi_c}|^2$ amplitudes each show two peaks, visible in Fig.~\ref{fig:J12}. They correspond to two poles of the coupled $B\Xi_c$--$B_s\Lambda_c$ channel in the present on-shell scheme. To address this point, we will employ the analysis done in~\cite{Xiao2019} to study the $P_{csb}$ system. The corresponding WT potential is given by
\begin{equation}
V=F\begin{pmatrix}-1 & -\sqrt2\\ -\sqrt2 & 0\end{pmatrix},\qquad F=\frac{p_i^0+p_j^0}{4f^2}.
\label{eq:block}
\end{equation}
The $B_s\Lambda_c$ channel has no diagonal force, since $\hat{\mu}_3=0$, so on its own it cannot bind. A controlled test, turning the off-diagonal coupling on and off, makes the mechanism explicit. With coupling off, the $B\Xi_c$ channel with diagonal $-F$ binds once just below its threshold at $7737.7$ MeV, and the $B_s\Lambda_c$ channel does not bind at all. The coupling raises the attractive eigenvalue of Eq.~\eqref{eq:block} from the single-channel $-F$ to $-2F$, whose eigenvector $(\sqrt{2}\,B\Xi_c+B_s\Lambda_c)/\sqrt{3}$ is dominated by $B\Xi_c$. This deepens the $B\Xi_c$ pole from $7737.7$ to $7717.8$ MeV, while the orthogonal eigenvector with the repulsive $+F$ eigenvalue does not bind. The energy dependence of the kernel then turns the single $-2F$ prediction into two zeros, producing a second lower pole at $7596.4$ MeV in the same sector. The lower peak is therefore generated dominantly by inter-channel dynamics, i.e., the channel with no diagonal WT attraction, $B_s\Lambda_c$, acquires a pole through its coupling to the attractive $B\Xi_c$ channel. The same pattern repeats by heavy-quark spin symmetry in the vector sector, where the $B^{*}\Xi_c$ state at $7766.5$ MeV is accompanied by a $B_s^{*}\Lambda_c$ partner near $7650$ MeV.

The double peak occurs exclusively in these two channels. 
It does not appear for the other four states. 
This behavior arises because only the $B^{*}\Xi_c$ channels couple to a lower, force-free $B_s^{*}\Lambda_c$ partner with an off-diagonal strength of $\sqrt{2}$. In contrast, the $B^{*}\Xi_c'$ and $B^{*}\Xi_c^{*}$ states occupy effectively single-channel diagonal $-F$ elements and bind only once. This result is consistent with the chiral unitary approach. Dynamically generated two-pole structures are a well-defined feature of coupled-channel WT dynamics. This phenomenon was first established for the $\Lambda(1405)$~\cite{Oller2001,Jido2003}. Later, it was recognized as a recurring feature dictated by the leading-order chiral interaction~\cite{Ji2023}. This includes the local hidden gauge description of $\Xi_b$ states, which closely aligned with the present framework~\cite{Yu2019b}. 
However, the mechanism in this work is distinct from the canonical $\Lambda(1405)$ case.  In that scenario, the two poles arise from two separately attractive $\mathrm{SU}(3)$ eigenchannels, namely the singlet and octet. In contrast, the potential in Eq.~\eqref{eq:block} possesses a single attractive eigenvalue of $-2F$. An energy-independent kernel would therefore yield only single pole. The second pole emerges because the WT interaction strength is energy-dependent, with $F\propto p^0(\sqrt{s})$. This explicit energy dependence renders the attractive eigenamplitude $1-2F(\sqrt{s})\,G(\sqrt{s})$ non-monotonic, allowing it to cross zero twice. We have verified this mechanism numerically. Turning off the off-diagonal coupling leaves a single $B\Xi_c$ pole. Furthermore, artificially degenerating the two thresholds keeps the two poles intact. This test confirms that the double peak is driven by the induced coupling and the energy dependence rather than by the threshold splitting.

It is worth noting that these two lower poles are excluded from the ten threshold-associated states listed in Table~\ref{tab:spectrum}. 
They are reported separately because they are more deeply bound relative to their dominated thresholds. In particular, the $B_s\Lambda_c$ pole appears at $7596.4$ MeV, which is $57$ MeV below its $7653.4$ MeV threshold, while its heavy-quark spin partner, the $B_s^{*}\Lambda_c$ pole, is located near $7650$ MeV, below the $7701.9$ MeV threshold. Within the coupled $B^{(*)}\Xi_c$--$B_s^{(*)}\Lambda_c$ system, both states exhibit a high dominated-channel compositeness of $X_{\rm dom}\simeq0.96$.

The lower pole is thus $96\%$ a $B_s\Lambda_c$ molecule in the dominated two-channel sector, with only an $5\%$ $B\Xi_c$ admixture. So, it is genuinely molecular. Nevertheless, we stress that this weight, like its absolute position, is more sensitive to the subtraction scheme and to the on-shell WT approximation than the near-threshold weights of Table~\ref{tab:spectrum}, because the pole lies far below its dominated threshold. 

This second state is prominent in the bottom sector but not in charm because the WT strength scales with the meson energy, $V\propto p^0\sim m_{\mathrm{meson}}$. Replacement $\bar{c}\to\bar{b}$ increases the relevant meson mass from $m_D\simeq1.9$ GeV to $m_B\simeq5.3$ GeV, a factor of approximately $2.8$. The same two-channel in the charm sector does possess a second zero, but with the weaker charm kernel it falls more than $200$ MeV below the $D_s\Lambda_c$ threshold, far outside the range where the on-shell linear WT extrapolation is reliable. This is why charm-sector studies do not report it. The stronger bottom kernel lifts this zero into the physically meaningful near-threshold region. We note a caveat. Ref.~\cite{Shen2022} reports only the upper states, the ones that we match to about $2$ MeV. The deep $B_s^{*}\Lambda_c$ partners, with binding of order $55$ MeV relative to their dominated thresholds, should therefore be quoted as an additional prediction of the strong coupled-channel regime, but their absolute positions carry a larger scheme uncertainty than the near-threshold poles in Table~\ref{tab:spectrum}.

\subsection{Experimental prospects}

Because the $udsc\bar{b}$ states carry five different quark flavors, they cannot decay to a light meson and a light baryon through the strong interaction, so the dominated two-body decay channels are the ones we have considered. The most accessible final states are $B_c\Lambda$ and $B_s^{*}\Lambda_c$. The high luminosity of the proton-proton collisions at LHCb makes it well suited for the direct production of these states, which could be searched for in the $pp\to B_c\Lambda X$ or $pp\to B_s^{*}\Lambda_c X$ processes as proposed in Ref.~\cite{Shen2022}. The five-flavor content provides a clean and highly distinguishable tag that should help separate these states from ordinary backgrounds.

\section{Conclusion}

We have predicted ten isoscalar molecular pentaquark states with the genuinely exotic five-flavor content $udsc\bar{b}$, spanning $7.72$ to $7.96$ GeV with $J^P=1/2^-$, $3/2^-$, $5/2^-$ and are all narrow. In addition, their compositeness is dominated by a single meson-baryon channel at the $0.98$--$0.99$ level with $\sum_i X_i\simeq1$, confirming a clear molecular structure. The spectrum follows from three embedded symmetries with a single adjustable parameter. The LHG symmetry determines the interaction coupling while HQSS combines the pseudoscalar-baryon and vector-baryon channels into a single coupled channel scenario, and HQFS relates the outcome to the hidden-charm strange sector where the LHCb $P_{cs}$ states are already present. This symmetry content is what distinguishes the present work. The ten pentaquark states come from the same coefficient matrices that describe the observed charm states, with only the hadron masses and one suppression factor changed, so they are a controlled extrapolation and not an independent model. The heavy-quark spin symmetry, which the earlier separate-sector study does not impose, predicts the spin multiplets and their near-degeneracies as a direct and testable signal, and it restores the $\Xi_c^{*}$ baryon, giving a $J=5/2$ state, a $B^{*}\Xi_c^{*}$ triplet and a $B\Xi_c^{*}$ state that the two-sector treatment cannot generate. Finally, the study of coupled channels reveals an additional binding mechanism that is much more effective in the bottom sector than in charm, i.e. $B_s^{*}\Lambda_c$ channels, which have no diagonal WT attraction, can generate poles through their coupling to $B^{*}\Xi_c$, because the WT interaction grows with the meson energy. All of these states offer a clean target for LHCb collaboration in the $B_c\Lambda$ and $B_s^{*}\Lambda_c$ invariant mass spectra.

\section*{Acknowledgments}
This research project is supported by National Research Council of Thailand (NRCT): (Contact No. N41A680287). DS is supported by Thailand NSRF via PMU-B [grant number B39G680009]. DS has also received funding support from the Fundamental Fund of Khon Kaen University.

\bibliographystyle{elsarticle-num}
\bibliography{refs_2_1}

\end{document}